\documentclass[showpacs,preprintnumbers,amsmath,prl,
graphicx,amssymb,superscriptaddress,twocolumn,nofootinbib]{revtex4}
\usepackage{graphicx}
\usepackage{dcolumn}
\usepackage{bm}

\renewcommand{\today}{\number\day\space\ifcase\month\or January\or 
 February\or March\or April\or May\or June\or July\or August\or 
 September\or October\or November\or December\fi\space\number\year}

\begin{document}
\title{Light WIMP Searches: The Effect of the Uncertainty \\  in Recoil 
Energy Scale and Quenching Factor}

\newcommand{\efi}{Enrico Fermi Institute, Kavli Institute for 
Cosmological Physics and Department of Physics,
University of Chicago, Chicago, IL 60637}

\affiliation{\efi}
														
\author{J.I.~Collar}\affiliation{\efi}

\begin{abstract}

Taking liquid xenon detectors as a case study, the importance of 
a robust recoil energy calibration 
as a prerequisite to a search for 
light-mass Weakly Interacting Massive Particles (WIMPs) is emphasized.  
Important shortfalls in the analysis of existing measurements of the relative 
scintillation efficiency ($\mathcal{L}_{\text{eff}}$) and ionization 
yield ($Q_{y}$) for nuclear 
recoils in liquid xenon are described, leading to the conclusion
that recent attempts to extract 
light-WIMP sensitivity limits from the XENON10 and XENON100 detectors are 
premature and overly optimistic. 

\end{abstract}

\maketitle


\section{I: Introduction}

Direct searches for dark matter WIMPs are remarkably difficult experiments 
where an attempt is made to extricate rare signals from backgrounds orders of 
magnitude more frequent. An additional complication arises from 
eventually having to 
demonstrate, if at all possible, that any irreducible events are more likely the result 
of WIMP interactions than a less exotic source. In the case of 
light-WIMPs below a mass of $\sim$10 GeV/c$^{2}$, the situation is markedly 
more complex, given the modest nuclear recoil energies expected from 
their interactions (few keV or 
even sub-keV, depending on WIMP and target masses), 
typically very close or below the energy threshold of the detectors. This 
is an energy region for which models of detector response 
and expected backgrounds can break down, 
background rejection becomes inadequate or non-existent, and events 
with a non-radioactive origin (microphonics, electronic noise) may be 
taken for a WIMP signal. Recent times have seen a renewed interest in 
this WIMP mass region. The drive has been the detection by the 
DAMA/LIBRA 
collaboration of an annual modulation in the low-energy counting rate 
of their scintillators, in agreement with expectations
from a dark matter galactic halo composed by light dark matter 
particles \cite{damaclaim}. A second source of interest comes from 
particle phenomenologies that generate dark matter 
candidates precisely in this mass range \cite{light}. The situation has been 
compounded by the observation of a so far unexplained background excess in 
CoGeNT \cite{cogent}, a detector targeted to search for light WIMPs, 
and more recently by an excess of low energy recoils in CRESST 
bolometers \cite{cresst,cresst2}, both in possible agreement with the 
DAMA/LIBRA results if the WIMP hypothesis is adopted \cite{danandco}.

The purpose of this note is to provide a critical inspection of the 
methods that have been proposed or employed to understand the 
response to low-energy nuclear recoils and
to derive a light-WIMP sensitivity from liquid xenon (LXe) devices. 
Important deficiencies are found in these methods. 
The conclusion  drawn is that the knowledge necessary to 
allow a reliable exploration of this WIMP mass 
region is presently absent for LXe-based detectors.

The outline of this paper is as follows: Section II deals with the 
difficulties in establishing a reliable energy scale for few 
keV$_{r}$ recoils 
in LXe, in particular via the so-called best-fit Monte Carlo method. 
Recent energy scales proposed using this method are compared to 
existing expectations and to independent measurements of ionization 
yield by low energy 
Xe$^{+}$  impact on surfaces, finding a large disagreement in both cases. 
Well-known physical processes presently neglected by 
this method are described. 
A commentary is provided on how their inclusion should affect the energy scale, 
easing the disagreements described. A brief attempt 
at the interpretation under different energy-scale 
scenarios of a low-energy feature present in a recently 
released XENON10 ionization spectrum is offered. Section III describes 
an important systematic effect neglected in the analysis of all 
XENON10 and XENON100 measurements of relative scintillation 
efficiency$^{1}$\footnotetext[1]{For the purpose 
of the present discussion $\mathcal{L}_{\text{eff}}$ can be 
considered the equivalent of the more broadly employed concept of ``quenching 
factor'', in this case the ratio between the yield of direct (S1) 
scintillation light produced by 
a nuclear recoil and that from an electron recoil of the same energy, 
at zero drift field.} ($\mathcal{L}_{\text{eff}}$) using monochromatic neutron scattering. A 
discussion of its impact on the energy dependence of 
$\mathcal{L}_{\text{eff}}$ is provided. Separately, the low 
quality of the experimental data in recent attempts by XENON100 to 
characterize $\mathcal{L}_{\text{eff}}$ is pointed 
out. Finally, these concerns are illustrated 
with data from ongoing quenching factor 
measurements at the University of Chicago. Section IV contains the 
conclusions. 

\section{II: Uncertainty in Recoil Energy Scale}

In a recent workshop presentation \cite{sorensenidm10}, the XENON10 
collaboration released a preliminary spectrum of events in their detector 
as a function of ionization yield, i.e., the number of 
electrons extracted from LXe via the application of an 
electric drift field $E_{d}$= 0.73 kV/cm. These electrons are further 
accelerated through the gaseous phase of the detector, producing a 
considerable VUV light emission via electroluminescence (the so-called S2 light), 
resulting in 
the registration of $\sim$25 photoelectrons in the photomultipliers
per each electron extracted from the liquid phase. 
The high gain provided by the electroluminescence
permits the detection of single electrons with good resolution. It 
is therefore expected that this ionization spectrum should reach down to a 
threshold in recoil energy of O(1) keV$_{r}$, allowing a search for light 
WIMPs. A dark matter search based exclusively 
on an analysis of the LXe ionization spectrum could in principle  
circumvent many recent discussions 
about the energy dependence
of the relative scintillation efficiency $\mathcal{L}_{\text{eff}}$, 
and how it impacts 
LXe sensitivity to light WIMPs \cite{juandan,reply}.
While electron-recoil background 
rejection is lost when considering just the ionization (S2 light) spectrum, the 
low-energy counting rate$^{2}$\footnotetext[2]{This background level ($\sim$0.06 ckkd 
at 20 keV$_{ee}$ before 
fiducialization, $\sim$0.01 ckkd following it), has been widely 
advertised  
in recent XENON100 presentations as being two orders of magnitude lower 
than in any competing dark matter detectors. In reality, it is comparable 
to the past generation of intrinsic germanium detectors ($\sim$0.03 ckkd, 
\cite{igex}).} achieved in 
XENON10 ($<$1 count / keV$_{r}$ kg day, 
Fig.\ 1) should be sufficient to investigate the remaining 
favored DAMA/LIBRA region \cite{danandco}, if the energy threshold is indeed as low as 
expected. However, an important requirement for the analysis of such 
ionization spectra in terms of sensitivity to light WIMPs is the 
establishment of a reliable correlation between the recoil energy 
scale and the ionization yield measured. During their recent 
presentation, XENON10 collaborators employed a new proposed recoil energy 
scale to generate the spectrum in Fig.\ 1 (top). If confirmed, it 
would amply exclude the WIMP phase-space regions presently compatible 
with a dark matter interpretation of DAMA/LIBRA, CoGeNT and recent preliminary 
CRESST data \cite{danandco}, at the very least for the standard 
present understanding of WIMP-nucleus interactions.

\begin{figure}
\includegraphics[width=7.7cm]{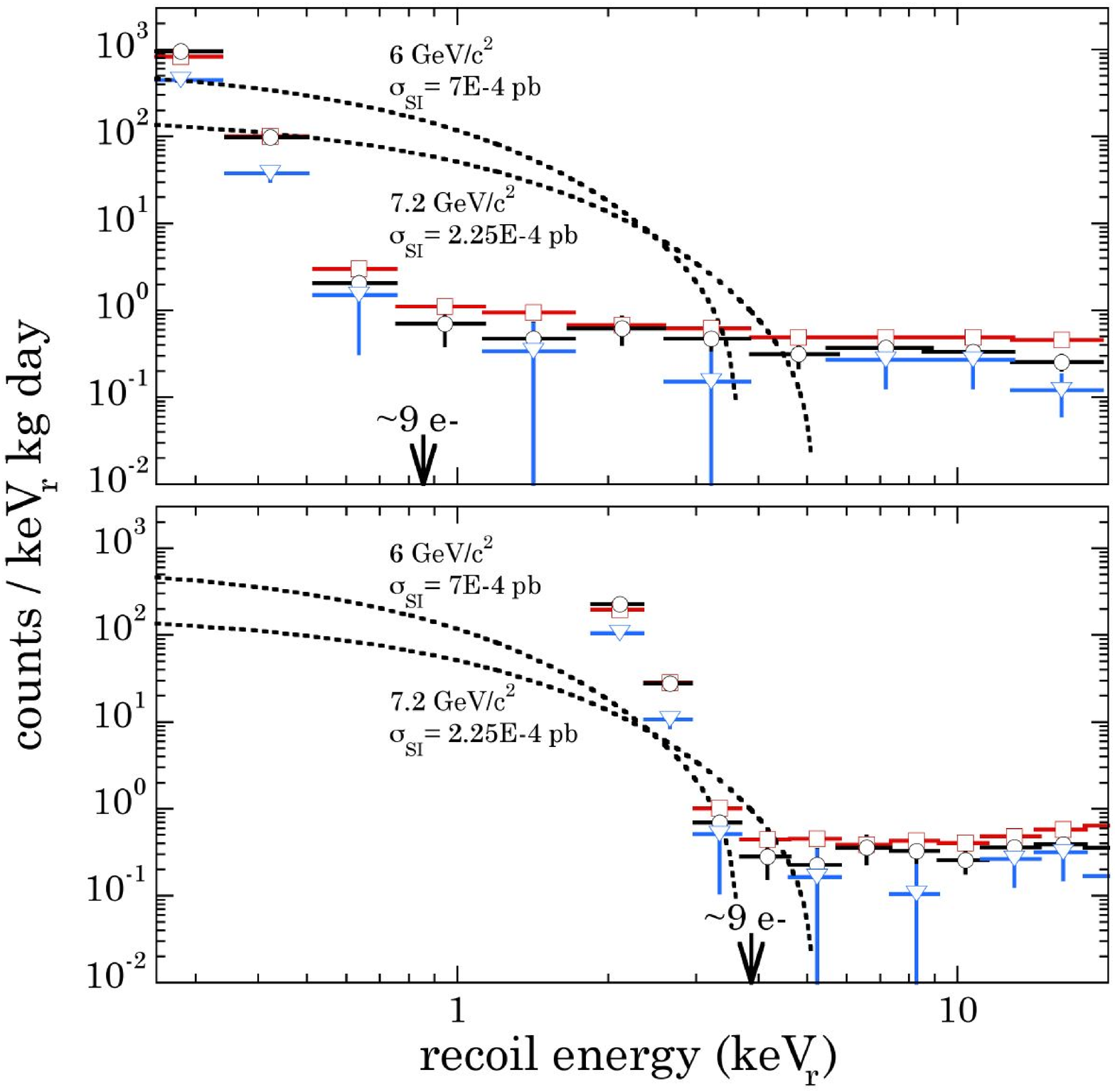}
\caption{{\it Top:} Spectrum of events in the XENON10 detector based on 
ionization yield only, as presented in \protect\cite{sorensenidm10}. 
The energy scale was defined by a ``best-fit Monte Carlo'' method. 
The original spectrum of events vs.\ electrons extracted 
from LXe can be found in \protect\cite{sorensenidm10}: the lowest 
bin shown here corresponds to an ionization yield of 3-4 electrons. 
The color code corresponds to different fiducial 
volume cuts \protect\cite{sorensenidm10}. The predicted signal for 
two example light WIMPs (dotted lines) is overlapped on the spectrum.
{\it Bottom:} Similar to top panel, but using the 
dotted expectation curves in Fig.\ 2 to define the recoil energy scale (see text). 
}
\end{figure}

Reservations were expressed in \cite{sorensenidm10} about the 
trustworthiness of the recoil energy scale employed to arrive at this 
spectrum. The vast extent to which this 
caution is advised is emphasized here, if the goal is to obtain reliable light-WIMP 
limits: 
Fig.\ 2 displays the most recent attempts 
\cite{sorensenidm10,xenonnim2009} by the 
XENON10 collaboration to establish a correlation between the 
ionization yield and the recoil energy scales, represented by solid color 
lines and matching color bands for the claimed one-sigma
uncertainties (statistical error only). The ionization yield vs.\ recoil energy is typically 
represented by the normalized quantity $Q_{y}$ (units of 
electrons/keV$_{r}$). For clarity, here  the actual 
measured quantity ($Q$, charge yield in number of electrons) is shown on the horizontal axis. 
All these energy calibrations originate in variations around a common 
method: a 
comparison is established at some point between experimental data obtained during  
exposure to a neutron source (commonly AmBe) and a Monte Carlo 
simulation of the expected spectrum of events vs.\ recoil energy. In 
the case of the latest attempt (\cite{sorensenidm10}, red line), this is accomplished by 
using a spline with mesh points fixed at arbitrary recoil energies  
(1,2,4,8\ldots 128,256 keV$_{r}$) to describe $Q_{y}$, 
allowing $Q_{y}$ to float unconstrained until the best possible match between 
the experimental ionization spectrum and the simulated spectrum of events 
vs.\ recoil energy is reached$^{3,4}$.\footnotetext[3]{No information 
was provided in \cite{sorensenidm10} on how discrepancies in the 
overall rate 
of simulated and measured events are dealt with during this 
procedure, or about the selection of fitting range. 
These choices can, on their own, alter the extracted energy scale.}\footnotetext[4]{A similar method has been used in the 
past to obtain $\mathcal{L}_{\text{eff}}$ \cite{peterphd}. At least 
there, the overall event rate normalization between simulation and 
data was treated as a free parameter, a highly questionable approach 
(see Sec.\ III) able to affect the low energy 
$\mathcal{L}_{\text{eff}}$ and $Q_{y}$.} The so obtained $Q_{y}$ defines the 
correlation between recoil energy and ionization 
scales.  As pointed out in \cite{juandan}, 
the original disagreement between simulations and LXe 
data is typically severe at low recoil energy: 
spectra of direct (S1) scintillation display a  
lack of response to AmBe neutron-induced recoils below few keV$_{r}$ when compared to the 
expectations, regardless of $\mathcal{L}_{\text{eff}}$ adopted \cite{theiraps}. 
Possible reasons for this are described 
below. While no mention of this initial state of affairs for 
the S2 light (ionization) spectrum 
was made in 
\cite{sorensenidm10}, the limited information provided indicates that 
this is the case too. The XENON10 collaboration refers 
to these methods collectively under 
the unnerving denomination ``best-fit Monte Carlo'' \cite{xenonnim2009}.

\begin{figure}
\includegraphics[width=7.5cm]{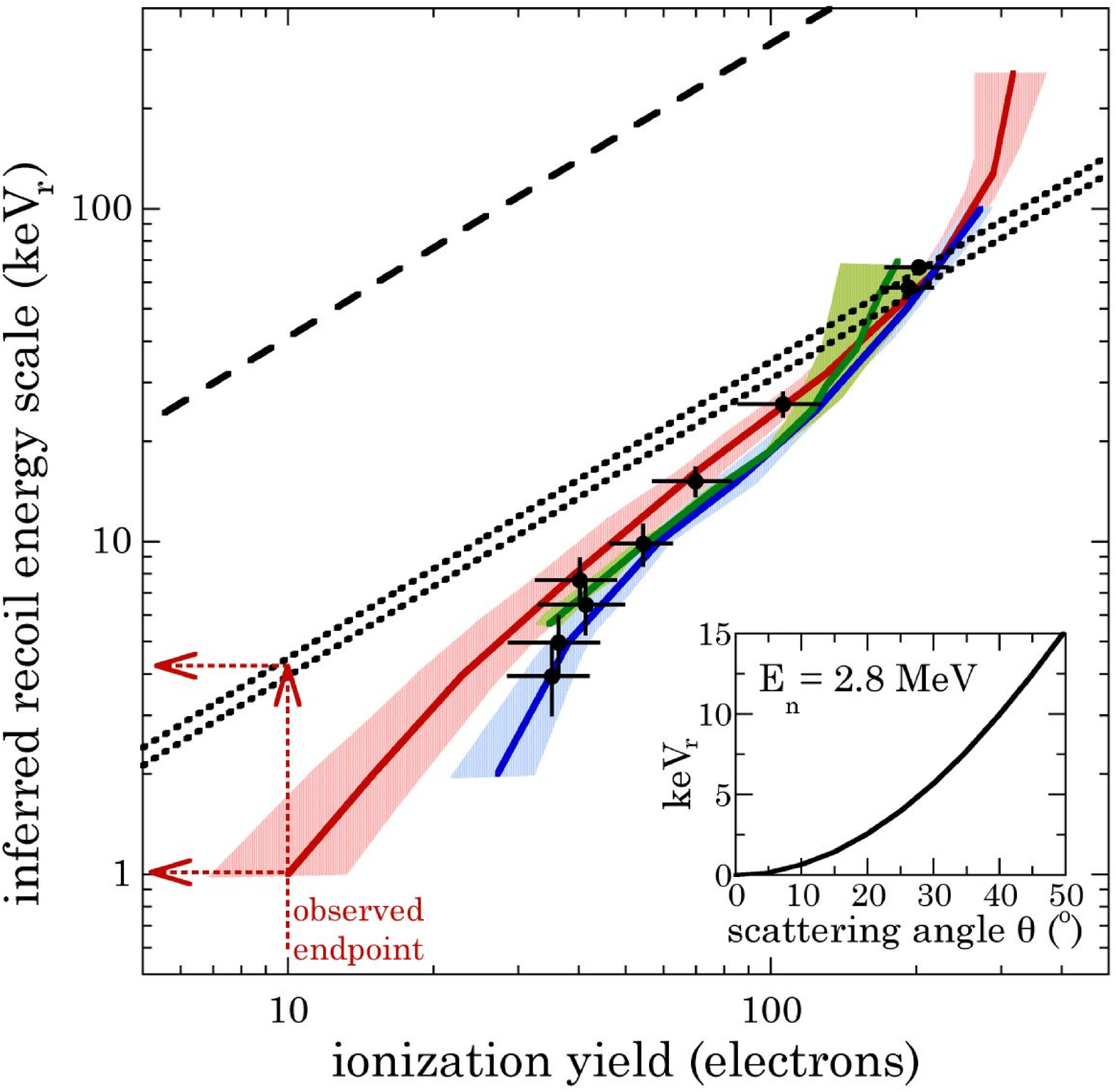}
\caption{Colored lines: correlations between recoil energy scale and 
ionization yield proposed by the XENON10 collaboration over the last 
two years, using best-fit Monte Carlo methods. 
The colored bands represent the claimed one-sigma 
statistical uncertainties (see text). Dotted lines correspond to the most 
recent expectations, based on the formalism presented in \protect\cite{prlexpec} (one based on
``Columbia'' detector data, the second on the ``Case'' detector). 
The dashed line indicates an 
earlier attempt to predict these \protect\cite{prevexp}. All 
expectations are for $E_{d}$= 0.73 kV/cm. The 
data points correspond to measurements by Manzur {\it et al.} 
\protect\cite{manzur,manzurphd}. Red arrows indicate the  
position of the endpoint for the low-energy rise in the spectra of Fig.\ 1  
(see text). {\it Inset}: correlation between 2.8 MeV neutron scattering angle and recoil energy 
deposited in LXe in the measurements by Manzur {\it et 
al.} \protect\cite{manzur}. 
}
\end{figure}

A reader familiar with the energy calibration of radiation detectors 
will readily identify the risks involved in this approach: all 
uncertainties in the inputs to the simulation, limitations to its 
accuracy$^{5}$,\footnotetext[5]{To this author's knowledge, Geant4 
(used in \cite{sorensenidm10}) has never been validated or 
verified in the $\sim$few keV to sub-keV recoil energy region, having 
historically exhibited difficulties in correctly modelling low-energy 
neutron transport. Several important improvements to the code in this respect 
are still in the making \cite{apostolakis}.} any experimental 
systematic effects, and most importantly, the effects of any 
physical processes not included in the simulation (such as the kinematic cutoff 
discussed below), will be 
automatically reflected as distortions
in the inferred recoil energy scale. In other words, in this method, the reasons for 
any initially 
noticeable disagreements between simulation and experimental data are not 
investigated, instead the recoil energy scale is used as a buffer to 
absorb these.  Taking as a reference a yield of 10 
electrons (the approximate endpoint for the observed low-energy rise 
in Fig.\ 1), the low-energy extrapolation of 
the colored curves in Fig.\ 2 indicates that this method of calibration results in 
recoil energy 
assignments for it anywhere in the $\sim$0-2 keV$_{r}$ 
interval.
No explanation is provided for the origin of the 
inflexion points and non-proportionality (particularly unrealistic at high 
energy) noticeable in those curves. 
The trend of a marked increase in $Q_{y}$ towards 
decreasing recoil energy in two of these curves (red, blue)
is against all expectations: as  
pointed out in \cite{juandan}, a kinematic cut-off at $\sim$40 keV$_{r}$ 
is expected for nuclear recoils in LXe, below which ionization should be adiabatically 
quenched. The same can be said of primary S1 scintillation, to the extent that it is 
expected to be mainly mediated by ionization \cite{juandan}. Indeed, the 
observed traditional deficit of low-energy recoil events in LXe under AmBe neutron irradiations mentioned 
above is easily understood under this consideration.

The origin for 
this kinematic cutoff is 
in simple two-body kinematics: when the largest possible energy 
imparted to a valence 
electron by a slow-moving recoiling ion falls below the minimum excitation 
energy of the system (a 9.3 eV bandgap in LXe), a progressive quenching of 
scintillation and ionization should be observed (other secondary processes such 
as for instance potential electron emission from Auger de-excitation in an 
ionized projectile, whenever allowed, or transient autoionizing quasimolecules can 
still play a role, leading to a smooth rather than sharp cutoff 
\cite{ahlen,ziegler,surface}). Contrary to 
an opinion recently expressed by the XENON100 collaboration \cite{reply}, the 
basic principles behind this intrinsic limitation are well-known, with abundant 
references in the experimental and theoretical 
literature on transport of slow ions \cite{ahlen,ziegler,surface}. This quenching 
has been observed at the predicted kinematic cutoff for hydrogen 
recoils in organic scintillator \cite{ahlen}. Preliminary evidence 
for it in Cs and I recoils in CsI[Na], also at the expected cutoff energy, is provided in Sec.\ III of this 
paper. Due to an unfavorable combination of nuclear mass and electron band-gap energy, 
LXe-based dark matter detectors should be particularly affected 
by this limitation: the cutoff for Na recoils in 
NaI[Tl] and Ge recoils in germanium crystals is predicted to appear below 
the threshold of present dark matter detectors  \cite{juandan}. 
Other less well-known radiation 
effects left out by the best-fit Monte Carlo method could be 
listed: for instance, as recoil energies become very small, 
a diminishing recoil 
track length relative to the range of ionized electrons 
may lower the chances of recombination, helping any 
charge still being generated to escape and contribute to the S2 light
\cite{manzur}. 

It cannot be overemphasized that none of these microscopic physical 
processes mediating the generation of information carriers 
(free electrons, primary scintillation photons) following a nuclear recoil are 
included in
popular simulation packages such as Geant4 or MCNP, which stop at 
generating a simple distribution of recoil 
energies. Just for this reason alone, the best-fit Monte Carlo 
method should be expected to generate extraneous structure in the 
correlation between recoil energy and ionization yield, aberrating
the true relationship between these two 
magnitudes$^{6}$.\footnotetext[6]{Leaving 
aside the limitations of the best-fit Monte Carlo method in its 
present form, the derivation of 
the new WIMP exclusion plot presented in \cite{sorensenidm10} begs 
justification: 
the one-sigma error 
bars on $Q_{y}$, converted here to the red band in 
Fig.\ 2, are exclusively statistical. They are therefore 
vastly underestimated, not including the uncertainties 
in the energy scale generated by this method. The broad dispersion in the 
curves depicted in Fig.\ 2 gives an idea of the magnitude of this 
(hard to quantify) uncertainty in the energy scale.
Even if this one-sigma band is accepted at face value, 
the customary WIMP exclusion contour should be extracted 
from a spectrum with recoil energy scale 
derived from a conservative excursion from a central 
$Q_{y}$ value, as allowed by 90\% C.L. uncertainties. 
Taking again the $\sim$10 electron endpoint as a reference and using 
Fig.\ 2 as a guide, it is possible to estimate that the 
90\% C.L. dark matter sensitivity claimed in \cite{sorensenidm10}
should be relaxed by more 
than an order of magnitude. To add to the confusion, a
markedly more conservative (by a factor $\sim$40) exclusion plot based on 
the XENON10 S2 spectrum was 
recently presented 
in \cite{peter}, using an identical $Q_{y}$, 
i.e., the same proposed recoil 
energy scale as in \cite{sorensenidm10}.} More specifically, an 
attempt to expand the method by including these processes
{\it prior} to the reconciliation of simulation and data should alter the 
behavior of the colored curves in Fig.\ 2: for instance, the adoption 
of an adiabatic term \cite{ahlen} to include the reduced 
efficiency for electronic excitation below kinematic cutoff 
should bring those curves closer to or beyond the expectations 
discussed next (an adiabatic term would rather naturally account 
for any observed decrease in 
response to AmBe neutron recoils at low-energy).
The best-fit Monte Carlo method, as it stands, cannot be 
accepted as a substitute for a genuine 
effort to understand these underlaying physical processes and the way 
they impact the recoil energy scale. 

During this last attempt at defining a recoil energy scale, 
the XENON10 collaboration neglected any mention to the subject of  
expectations. These exist, and have been discussed by this 
collaboration
and other workers before. The dotted lines in 
Fig.\ 2 
show the expected correlation between recoil energy E$_{rec}$ and ionization yield, 
calculated following the relationship described in a recent XENON10 
publication \cite{prlexpec}, $Q(E_{d})=0.2~ \mathcal{L} 
E_{rec}/W_{e}$, where $\mathcal{L}$
is the Lindhard nuclear quenching factor$^{7}$\footnotetext[7]{A 
discussion of the 
relationship between $\mathcal{L}_{\text{eff}}$ and $\mathcal{L}$ 
(the second sometimes referred to as $q_{ncl}$) can be found in 
\cite{manzur}.} (extracted here from 
SRIM2010 \cite{SRIM2010}), $W_{e}$= 15.6 eV is the average energy spent by an 
electron recoil to form an ion-electron 
pair \cite{W}, and the charge yield at a drift field $E_{d}$ relative to that 
at infinite field $Q(E_{d})/Q(\infty)=0.2$ 
was measured at 56.5 keV$_{r}$ in \cite{prlexpec}. These 
expectations, based on the Lindhard-Scharff-Schiott theory through 
their dependence on $\mathcal{L}$, 
generate a $Q_{y}$ roughly constant in energy, i.e., a  
proportionality between recoil energy and ionization. It must be kept 
in mind that the effects of the kinematic cutoff may not be 
completely accounted for in such models, as noticed experimentally in \cite{ahlen}. In 
other words, more sophisticated models generating expectations 
tending to the horizontal at few keV$_{r}$ in Fig.\ 2  can be 
constructed. 

There is an evident tension between these expectations and the ever 
mutable low-energy scales so far 
proposed by XENON10, most certainly the result of the 
artificial structure that the present best-fit Monte Carlo method is expected 
to introduce. However, it should be kept in mind that 
this expectation is itself subject to 
considerable uncertainty, and therefore should be regarded as  
simple guidance. For 
instance, the value of $Q(E_{d})/Q(\infty)=0.2$ employed to 
generate the dotted curves in Fig.\ 2 
may have an unknown variation over the broad energy range depicted in this 
figure.  Similarly, while the
latest version of SRIM \cite{SRIM2010}, used here to generate 
$\mathcal{L}$, offers 
a considerably better agreement with existing LXe measurements than
previous ones \cite{srimdis}, it should be kept in mind that 
SRIM predictions are semiempirical, i.e., biased by low-energy quenching factor measurements 
that may in turn be flawed. In particular, an absolute cutoff for $\mathcal{L}_{\text{eff}}$
at few keV$_{r}$, a possibility strongly argued for in Sec.\ III, 
would suggest an expectation curve in Fig.\ 2 tending to the horizontal 
at this cutoff energy. 

Just to 
illustrate the incipiency of the knowledge about the processes 
mediating the generation of 
ionization and scintillation by low-energy recoils in LXe, 
a dashed line in Fig.\ 2 shows the expectations for nuclear recoils 
put forward by 
XENON10 as recently as in \cite{prevexp}, which were inferred from
observations made using 
alpha particles.  

Besides the colored curves in Fig.\ 2, generated via the best-fit 
Monte Carlo method, few other measurements of $Q_{y}$ exist. 
Measurements of $Q_{y}$ in \cite{prlexpec}, not shown in Fig.\ 2 for clarity, 
approximately follow the blue curve over the range 20-100 keV$_{r}$. 
However, the recoil energy scale used  
to extract $Q_{y}$ in \cite{prlexpec} was derived using a debatable choice of
$\mathcal{L}_{\text{eff}}$. The methodology of the calibrations that generated 
this  $\mathcal{L}_{\text{eff}}$ curve  will be 
sharply criticized in Sec.\ III of this paper. A rapidly decreasing  
$\mathcal{L}_{\text{eff}}$ towards zero recoil energy
like the alternative proposed in Sec.\ III would result in changes to the 
recoil energy scale that would make
the $Q_{y}$ derived in \cite{prlexpec} tend towards the 
dotted expectation lines rather than the blue curve in Fig.\ 2. 
Erroneous conclusions from one LXe calibration can rapidly propagate 
to another, such is the complexity of the relationship between 
quantities and physical processes in LXe detectors. 

The data points shown in Fig.\ 2 deserve special attention. 
They correspond to measurements by Manzur {\it et al.} \cite{manzur} 
at a value of $E_{d}$ similar to that used in 
the XENON10 detector, with the important difference that their energy 
scale is in principle
known, independently determined by the kinematics of the elastic scattering of 
monochromatic 2.8 MeV neutrons from a xenon target. The apparent trend in those 
measurements is one of an increasing $Q_{y}$ with decreasing recoil 
energy. Not only this clashes with the expected behavior below 
kinematic cutoff,
but it also poses an interesting number of questions: the low-energy 
extrapolation of these datapoints in Fig.\ 2, if confirmed, 
would imply that the low-energy rise below $\sim$10 electrons observed in the 
XENON10 ionization spectrum (Fig.\ 1) cannot originate in nuclear 
recoils. An alternative would be a process not the immediate result of 
a particle interaction, for instance, an ``spontaneous'' multiple electron emission.
Spontaneous single-electron emission from LXe 
is known to follow large energy depositions by gamma 
rays, with a half-life of 
$\sim$100 $\mu$s, and is possibly due to photoionization \cite{spont}. 
Delayed emissions are not uncommon in 
some detecting media: for instance, a few inorganic scintillators 
(e.g., CsI[Tl]) are 
notorious for an ``afterglow'' from long-lived phosphorescent 
states, involving the release of single scintillation photons. 
However, it is very hard to envision an atomic energy-storage 
mechanism in LXe that 
would lead to the delayed emission of up to $\sim$10 electrons. 
The high gain afforded by the electroluminescence clearly points at a 
multiple rather than single electron emission. The alternative left 
is to ascribe an origin in minimum ionizing particle interactions
to this observed rise in rate below 10 electrons. The value of $Q(E_{d})/Q(\infty)$ 
measured in \cite{prlexpec} for electron recoils suggests that this 
rise would then have an endpoint at an electron-equivalent energy of 
a mere 
$\sim$0.25 keV$_{ee}$. Compton scattering from gammas is not expected to 
generate any such features (the Klein-Nishina relation does not 
include a highly forward-peaked excess). Beta decay or 
Bremsstrahlung interpretations concentrated this low in energy are not 
plausible. 
The chance of this signal arising 
from degraded surface activity seems to be excluded by its survival 
under different fiducial volume cuts (Fig.\ 1, \cite{sorensenidm10}). 
No other meaningful processes come readily 
to mind. While an understanding of this low-energy feature would be 
desirable, the apparent low-energy trend in the $Q_{y}$ measurements by Manzur 
{\it et al.} limits the interpretations.

The  measurements of $Q_{y}$ by Manzur {\it et al.}, taken at face value, generate a third 
concern. The processes governing 
electron emission by slow ions immediately prior and during 
impact on surfaces are not identical to 
those expected from recoils in bulk LXe, but the dominant mechanisms are in common. 
Therefore, it is educational to contrast the several tens of 
electrons generated by a 1 keV$_{r}$ Xe recoil that would be implied by the 
extrapolated low energy $Q$ postulated by Manzur {\it et al.}, with the observed 
electron yields of O(10$^{-3}$) e$^{-}$/ion generated by a Xe$^{+}$ 
ion impact of the same energy on W or Au electrodes \cite{surface}. In this energy regime 
the ion range and the mean escape depth of electrons from the 
electrode are 
comparable (few nm, \cite{surface}), i.e., the measured yield is representative of the 
actual total ionization generated. This very small yield is of the same order of 
magnitude of 
what would be predicted by the dotted expectation line in Fig.\ 2 
after the 
introduction of an adiabatic term \cite{ahlen} to account for the kinematic cutoff. 

In trying to find a solution to these wee conundrums, this author 
searched \cite{manzur,manzurphd} for a description of the 
method of alignment between 2.8 MeV neutron source, LXe cell, and 
scintillator cell used to detect scattered neutrons (the scattering 
angle defining the recoil energy in the measurements). None was found, and also an absence of 
treatment for the uncertainty 
in recoil energy that any angular misalignment would bring, 
therefore unrealistically assumed to be nil$^{8}$.\footnotetext[8]{As 
it turns out, no goniometric equipment was employed by Manzur {\it et al.} The 
approximate 
scattering angles were derived from photographs of the setup taken 
from the vertical \cite{priv}. 
In the opinion of this author, 
this method does not guarantee the desired angular accuracy discussed 
in the main 
text.} Based on 
the geometry of this experiment, size of the detector cells and the 
relatively short distance between them, 
a modest 
misalignment in the relative position of the cells and source
could enlarge the recoil 
energy error bars significantly and/or shift the datapoints in Fig.\ 2 in energy:
as can be seen from the inset in Fig.\ 2, a cumulative misalignment by 
as little as $5^{\circ}$, which amounts in the experiment by 
Manzur {\it et al.}
to 2-3 cm of 
cumulative displacement in the 
assumed relative position between the detector cells and/or between 
them and the neutron source, is 
sufficient to remove the paradoxical situations described 
above (e.g., the lowest energy datapoint in Fig.\ 2 shifts from 4 keV$_{r}$ to 
6 keV$_{r}$ over $5^{\circ}$). A certain discomfort arises from 
the simultaneous realization that the XENON10 
light-WIMP sensitivity claimed in \cite{x10leff} using the 
$\mathcal{L}_{\text{eff}}$ measurements from Manzur {\it et al.}
is also critically dependent on this matter: the same unaccounted-for
displacements by few cm would result, quite literally, in a decrease in XENON10 
light-WIMP sensitivity by orders of magnitude. While this 
should be reason enough to take the limits claimed in \cite{x10leff} 
with caution, 
much graver concerns about the methodology used by Manzur 
{\it et al.} and earlier authors to arrive to 
$\mathcal{L}_{\text{eff}}$ and $Q_{y}$ will be expressed in Sec.\ III.
Considerably less information (trigger and software efficiency, S2 
gain, analysis procedure, etc.) is 
provided by Manzur {\it et al.} in 
\cite{manzur,manzurphd} on the subject of their derivation of $Q_{y}$, 
as compared to their very extensive treatment of 
$\mathcal{L}_{\text{eff}}$, preventing an additional 
investigation here of how the neglected systematic effect 
described in Sec.\ III 
may have affected the $Q_{y}$ datapoints in Fig.\ 2. A skewness 
similar to that affecting the low-energy S1 light peaks 
discussed in Sec.\ III can be noticed in the single S2 
light example shown by Manzur {\it et al.} (Fig.\ 8c in \cite{manzur}), 
possibly indicating the presence of S2 threshold effects at the level of 
$Q\sim$few electrons. The reader is referred to Sec.\ III for more 
details. 

As a last remark on the difficulties in the determination of  
light-WIMP sensitivity via S2 light, the bottom panel in Fig.\ 1 
shows the effect of adopting the {\it expected} energy scale (central value of dotted 
lines in Fig.\ 2) rather than that generated by the latest version of 
the best-fit Monte 
Carlo, taking into account the changes in bin width that the 
energy translation results in.  It would be possible to obtain a 
near-perfect agreement between the example light WIMPs in the figure 
(intentionally chosen in the region of interest 
for DAMA, CoGeNT and CRESST \cite{danandco}) and this 
spectrum, with the  introduction of the adiabatic term to account for the 
expected kinematic cutoff or by examining other uncertainties in the 
expectations. It is sobering to realize that one can go from this 
bizarre coincidence to 
severe exclusion of the candidates
over a shift in a notoriously ill-defined energy scale by a 
mere $\sim 3$ keV$_{r}$. While the presence of a high-rate, 
homogeneously-distributed signal at very 
low-energy in the XENON10 or XENON100 detector would be hard to explain 
in an efficient self-shielding medium like LXe without invoking 
weakly-interacting particles, any such possible agreement with these other 
detectors should be presently
de-emphasized: at the time of this writing, any 
low energy scale for LXe should be considered highly speculative, 
such is the paucity of reliable experimental data
and detailed models of recoil response available for LXe.

\section{III: Uncertainty in Quenching Factor}

The information presently available on  
scintillation and ionization yield from low-energy nuclear 
recoils in LXe is plagued by contradictory observations and 
interpretations. In particular, 
$\mathcal{L}_{\text{eff}}$ is expected to decrease with decreasing 
recoil energy \cite{juandan} for the reasons given above, but 
published measurements display all possible behaviors. Recent attempts 
to reconcile these \cite{mana} fall short, lacking in a
critical examination of the analysis methods and experimental 
techniques utilized in the existing calibrations. In order to 
elucidate some of these issues, extensive MCNP-PoliMi \cite{polimi} 
simulations of the most recent assessment of 
$\mathcal{L}_{\text{eff}}$, by Manzur {\it et al.} in 
\cite{manzur,manzurphd}, have been performed here. This 
examination has revealed 
several issues affecting this type of 
measurement$^{9}$.\footnotetext[9]{Several of these issues are minor, 
and yet illustrative of the lack of scrutiny with which some of these results 
have been embraced. For instance, simple geometrical considerations 
suffice to demonstrate that the distance between LXe and 
scintillator cells quoted in \cite{manzur} (16-20 cm, information not mentioned in 
\cite{manzurphd}) would lead to 
an uncertainty in the recoil energies probed almost three times 
larger than claimed. This was confirmed by initial simulations. 
This issue is traceable to a mistake in   
units of distance in \cite{manzur}, where inches should have been used \cite{priv}.} 
Attention was paid to follow the prescriptions 
given in \cite{manzur,manzurphd} regarding energy resolution, S1 
light yield at different drift field values, cuts affecting 
the data, 
effect of threshold efficiency, etc., obtaining a good agreement with the 
particulars provided in \cite{manzur,manzurphd} on distribution of 
recoil energies, contributions from different types of events 
(inelastic scatters, multiple scatters, scattering on inert materials, etc.), and neutron
time-of-flight (TOF)
information. In this respect, Fig.\ 3 here allows a direct comparison to Fig.\ 
9 in \cite{manzur}. 

\begin{figure}
\includegraphics[width=7.7cm]{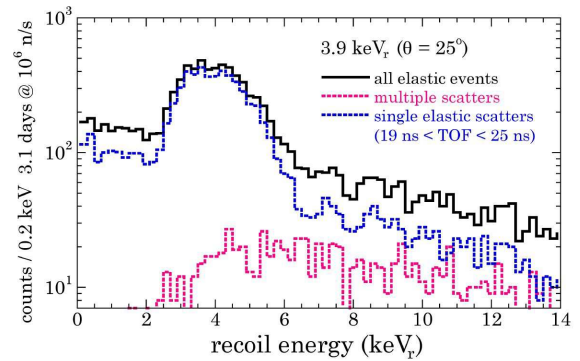}
\caption{MCNP-PoliMi simulated distribution of recoil energies from different components 
of the spectrum measured by Manzur {\it et al.}, 
comparable to Fig.\ 9 (top) in \protect\cite{manzur}. The log scale
generates subtle differences in appearance between these figures. }
\end{figure}

An important limitation to the measurements by Manzur {\it et al.} is 
not readily appreciated from a cursory inspection of 
\cite{manzur}: 
the position of the peak or 
maximum in the measured distributions of S1 scintillation 
light for neutron scattering angles below 
$\theta\sim60^{\circ}$ is entirely defined by the effect of the threshold 
efficiency (software and triggering). In other words, for 
recoil energies below $\sim$20 keV$_{r}$, the 
left shoulders responsible for the noticeable skewness in
distributions of S1 light like those in 
Fig.\ 8a in \cite{manzur} and Fig.\ 6.3 in \cite{manzurphd}
are solely the result of this threshold efficiency: the 
true scintillation maximum is expected at  
values much lower than the position of this peak. These true maxima 
are not reachable due to an scintillation yield 
(4.3 to 10.8 photoelectrons per keV of electron equivalent 
energy, PE/keV$_{ee}$, depending on drift field and mode of 
operation) that is insufficient for an optimal exploration of such low recoil 
energies. For 
instance, assuming a value of $\mathcal{L}_{\text{eff}}=$0.1 at 6 
keV$_{r}$, this peak would then be expected at between 2.6 and 
6.5 PE, whereas the effect of the threshold 
efficiency starts to be 
noticeable already at $\sim$15 PE for 
both the single 
and two-phase modes of operation. The threshold efficiency is provided 
in Fig.\ 10 in \cite{manzur} and top of Fig.\ 4 here. 

\begin{figure}
\includegraphics[width=7.7cm]{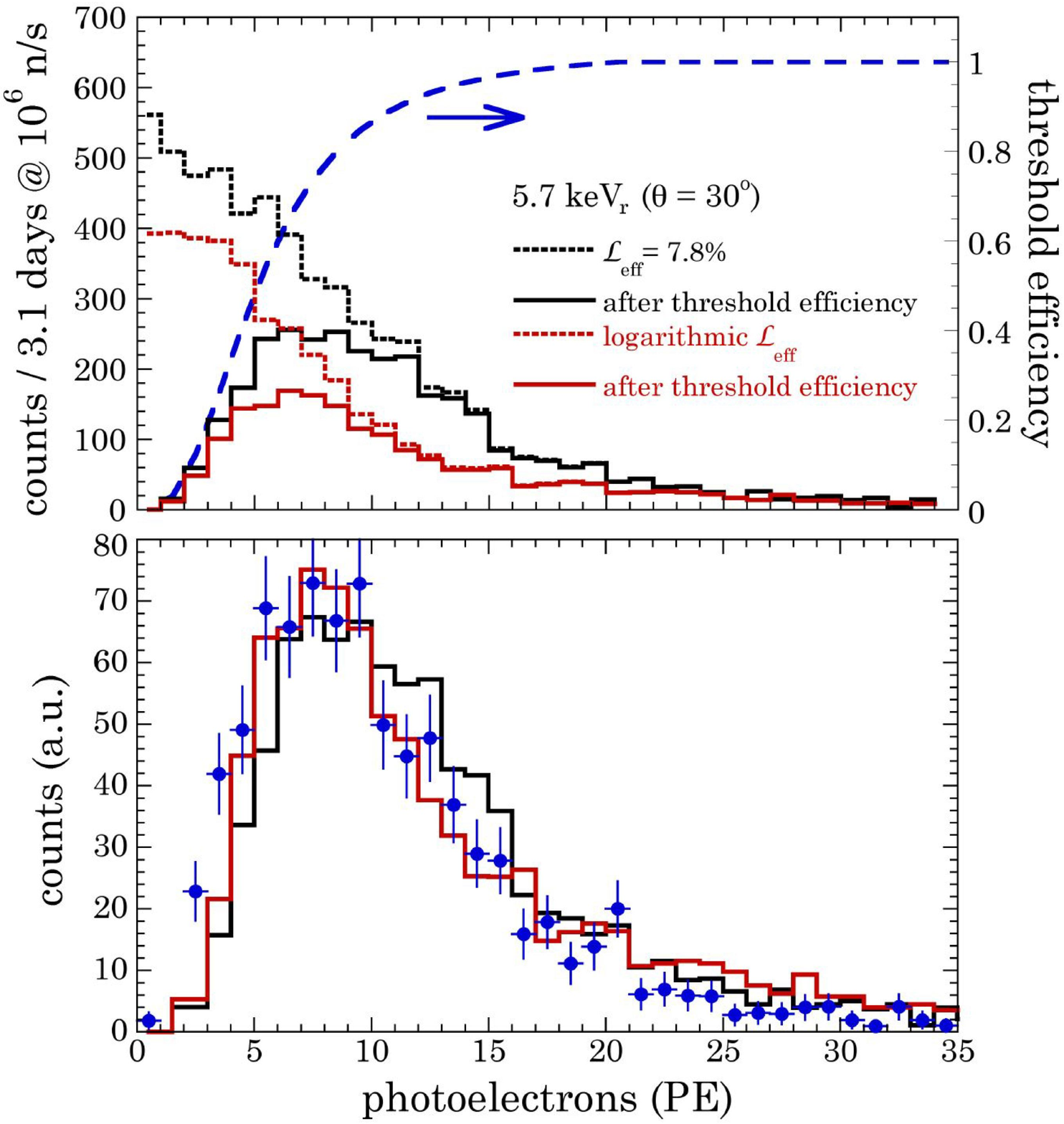}
\caption{{\it Top:} Simulated distributions of S1 light in the 
experimental setup of Manzur {\it et al.} 
\protect\cite{manzur,manzurphd}, for 5.7 
keV$_{r}$ recoils. Black histograms use the nominal 
$\mathcal{L}_{\text{eff}}$ claimed in 
\protect\cite{manzur,manzurphd} for this recoil energy, red histograms a  
logarithmic fit to their datapoints above 20 keV$_{r}$ (dashed line in 
Fig.\ 5 bottom, see text). The effect of the threshold efficiency 
is to shift the peak or maximum in the distributions to an 
artificially larger 
value. Both $\mathcal{L}_{\text{eff}}$ hypothesis 
remain distinguishable through their predicted event rate (area 
under solid 
histograms). {\it Bottom:}  The ability to distinguish these two largely different 
values of $\mathcal{L}_{\text{eff}}$ is lost once a normalization to the experimental 
rate (data 
points
from Fig.\ 8a in \protect\cite{manzur}) is performed, an unfortunate 
method of analysis
followed by Manzur {\it et al.} and earlier authors 
\protect\cite{prevexp,manamana1,manamana2}.}
\end{figure}

\begin{figure}
\includegraphics[width=7.7cm]{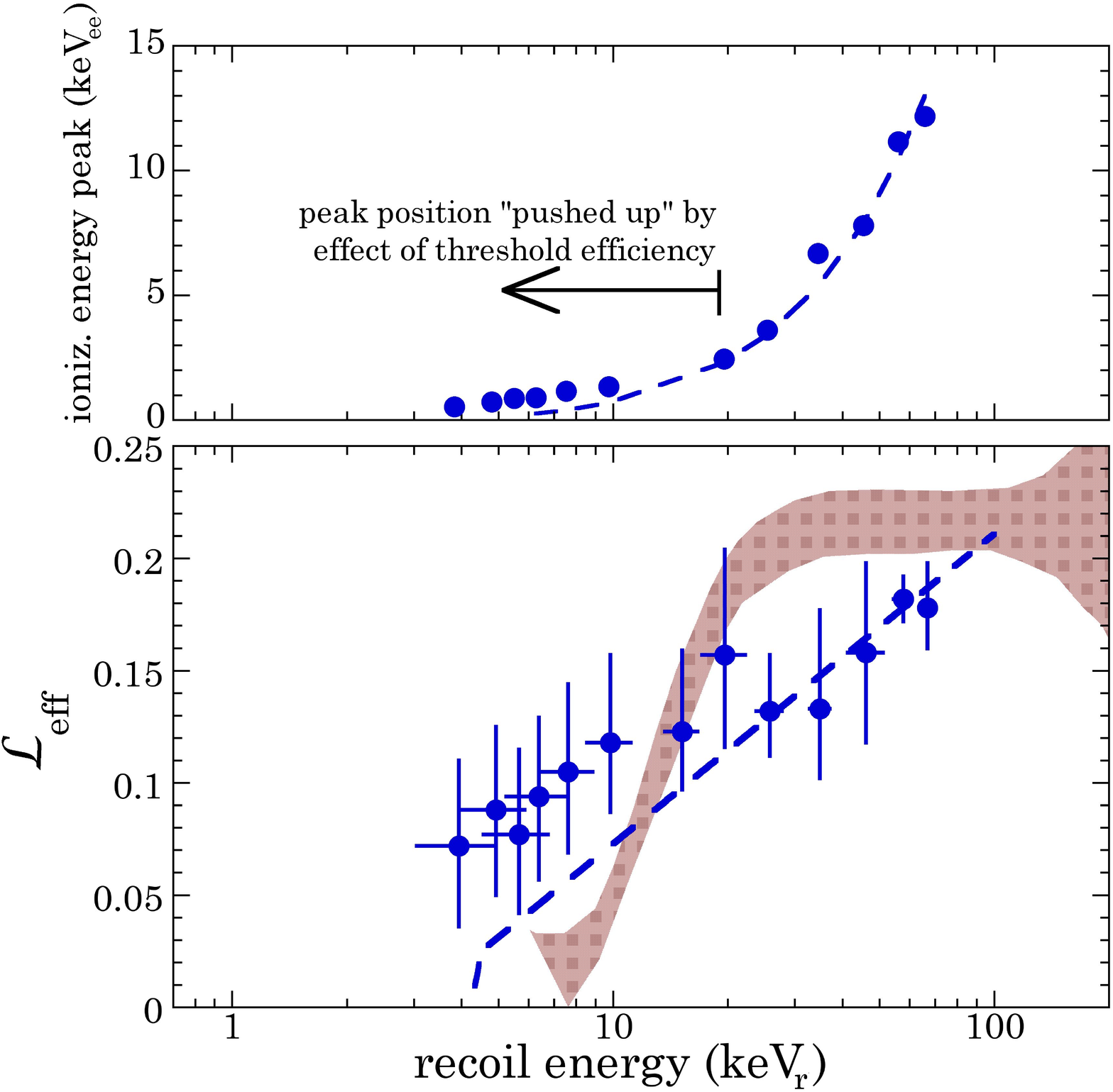}
\caption{{\it Top:} Position of the peak or maximum in S1 
scintillation (units of electron equivalent energy) observed by 
Manzur {\it et al.} as a function of recoil energy \protect\cite{wong}. The dashed line 
is derived from the logarithmic fit in the bottom panel, under the 
premise that the ratio of peak keV$_{ee}$ to keV$_{r}$ should 
be equal to $\mathcal{L}_{\text{eff}}$ in ideal experimental conditions (see 
text). {\it Bottom:} $\mathcal{L}_{\text{eff}}$ datapoints from 
Manzur {\it et al.} together with the $\mathcal{L}_{\text{eff}}$ derived 
by the ZEPLIN collaboration (red band \protect\cite{zeplin2}). The 
dashed line is a logarithmic fit to datapoints not affected by 
threshold efficiency ($>20$keV$_{r}$). Lower energy 
datapoints deviate from the fit by the effect of the threshold 
efficiency (see text). Cutoffs to this fit anywhere in the range 
3-6 keV$_{r}$ produce fits to the experimental data similar in 
quality to those 
shown in the bottom panel of Fig.\ 4. }
\end{figure}

This is not an insurmountable limitation, as long as careful 
attention is paid to the comparison between Monte Carlo simulations 
and data that ultimately leads to a best-fit 
$\mathcal{L}_{\text{eff}}$ for each scattering angle. Unfortunately, 
in the experiments described by Manzur {\it et al.} no attempt was made 
to include a comparison between the expected recoil rate and that 
observed$^{10}$.\footnotetext[10]{This was prevented by difficulties in 
controlling the stability of the DD neutron generator output towards 
the end of its target life,
and by throughput limitations of the data acquisition system during 
runs at the 
smallest scattering angles \cite{priv}.}
Instead, the simulated distributions of S1 light yield were 
normalized to the observed event rate prior to the chi-square analysis 
leading to the best-fit value for 
$\mathcal{L}_{\text{eff}}$, a step mentioned in  
\cite{manzur} and described with more detail in Sec. 6.\ 1.\ 2 of 
\cite{manzurphd}.

The nature of the problem created by this form of data analysis is  
illustrated in Fig.\ 4, 
using the 6 keV$_{r}$ S1 single-phase measurement provided 
as an example in \cite{manzur}. The top panel 
displays the expected S1 light yield distribution for the nominal 
$\mathcal{L}_{\text{eff}}=$0.078 obtained by Manzur {\it et al.} at 
this energy$^{11}$,\footnotetext[11]{$\mathcal{L}_{\text{eff}}$ is 
expected to be an energy-dependent quantity. Use of a constant 
value across the measured S1 spectrum during the chi-square analysis, 
as done in \cite{manzur,manzurphd}
is less than ideal. An effort was nonetheless made in \cite{manzurphd} to 
examine the effect of a linear energy dependence around the best-fit $\mathcal{L}_{\text{eff}}$.}
derived from the present simulation following the 
data cuts and energy resolution prescribed in \cite{manzurphd}, before 
(dotted histograms) and after (solid histograms) including 
the effect of threshold efficiency. Its equivalent, but 
using a considerably smaller and energy-dependent $\mathcal{L}_{\text{eff}}$
derived from a logarithmic fit (Fig.\ 5, bottom) to $\mathcal{L}_{\text{eff}}$ 
datapoints with E$_{rec}>$20 keV$_{r}$ (i.e., those not affected 
by the effect of threshold efficiency) is also displayed. As can 
be observed from the solid histograms in the top panel of Fig.\ 4, these two possibilities 
lead to distinguishable predictions, specifically a decrease in recoil rate per 
unit exposure to the neutron source by 40\% in the second scenario. 
However, once the normalization
to the experimental rate is performed (Fig.\ 4, bottom), this information is lost and
both $\mathcal{L}_{\text{eff}}$ scenarios yield a comparable good fit to 
the experimental data. In conclusion, the method of analysis followed in 
\cite{manzur,manzurphd} (and earlier experiments discussed next) cannot 
distinguish between these two largely different values of 
$\mathcal{L}_{\text{eff}}$.

Useful additional information was presented in an unpublished progress 
report \cite{wong} and reproduced here in 
the top panel of Fig.\ 5, where the position of the observed peak or maximum 
in S1 light yield (in units of keV$_{ee}$ rather than PE) is 
displayed as a function of mean recoil energy for early measurements by 
Manzur {\it et al.} For data points 
not affected by the threshold efficiency ($>$20 keV$_{r}$)
the ratio between both energies is, as expected by the definition of $\mathcal{L}_{\text{eff}}$, 
close to the value 
of $\mathcal{L}_{\text{eff}}$ eventually obtained (Fig.\ 5, bottom). 
For smaller recoil energies, for which the 
threshold efficiency affects and defines the position of the peak, this 
approximation breaks down, as a result of the true position of this peak 
having been artificially ``pushed up'' in energy, the effect 
illustrated here in Fig.\ 4 (top). 
This can be observed in, for instance, 
Figs.\ 11 a,c in \cite{manzur}, where the chi-square minimum 
 appears at $\mathcal{L}_{\text{eff}}\sim$0.055, rather than at the 
larger $\sim$0.12 that would be naively expected from said ratio and a 10.8 
PE/keV$_{ee}$ light yield. This trend in \cite{manzur,manzurphd} for the chi-square minimum
to appear well-below an expectation based on the (artificial) 
position of the 
scintillation peak, is in 
itself pointing at a rapidly decreasing $\mathcal{L}_{\text{eff}}$ towards zero recoil 
energy. But this is recognized in \cite{manzur,manzurphd} 
and not the point at stake: instead, the top panel in Fig.\ 
5 reveals that the measurements by Manzur {\it et al.} not affected 
by the threshold efficiency may very well be pointing at a vanishing
$\mathcal{L}_{\text{eff}}$ somewhere in the few keV$_{r}$ region, a behavior 
not dissimilar to that claimed by the 
ZEPLIN collaboration \cite{zeplin2}. 
The change in slope in Fig.\ 5 (top) appears {\it exactly} where 
expected
from the limitations imposed by threshold efficiency and light yield in this 
experiment$^{12}$.\footnotetext[12]{The asymmetry in the S1 distribution 
due to the effect of threshold efficiency is already clearly visible in the 10 
keV$_{r}$ data in Fig.\ 6.3 of \cite{manzurphd}. At 20 keV$_{r}$ 
and assuming a $\mathcal{L}_{\text{eff}}\sim$0.1-0.15, 
at least the lower light yield, non-zero drift field measurements are 
expected to be affected (the 
$\mathcal{L}_{\text{eff}}$ datapoints of Manzur {\it et al.} are an 
average of all conditions tested for each 
angle).}
It is inevitable to suspect that the 
combination of insufficient light yield and regrettable 
normalization of simulated to experimental 
rates used by Manzur {\it et al.} and others \cite{prevexp,manamana1,manamana2}
may have biased $\mathcal{L}_{\text{eff}}$ 
towards considerably 
larger values than those to be measured in more ideal 
conditions, or at least when properly taking into account systematic 
differences between expected and observed signal 
rates.
The logarithmic fit to the points above 20 keV$_{r}$ employed to 
illustrate 
this discussion (Fig.\ 5, bottom) lays slightly beyond the one-sigma 
uncertainty in $\mathcal{L}_{\text{eff}}$ quoted by Manzur {\it et 
al.}$^{13}$\footnotetext[13]{An 
unquantified very large
increase in signal rate towards small scattering angles was noticed in the 
measurements by Manzur {\it et al.} \cite{priv}. Present simulations 
predict that this rate should 
have increased by a factor $\sim$200 in going from 
$\theta=$125$^{\circ}$ (67 keV$_{r}$) to 
$\theta=$25$^{\circ}$ (4 keV$_{r}$), before accounting for 
threshold effects. Yet only a small deficit ($\sim$40\%, Fig.\ 4 top 
panel) in rate is 
expected from an $\mathcal{L}_{\text{eff}}$ much smaller than 
that determined by Manzur {\it et al.}, leaving ample 
margin for a severe misidentification of the low-energy
$\mathcal{L}_{\text{eff}}$, even more dramatic than what has been 
contemplated here.} The adoption of a 
similar $\mathcal{L}_{\text{eff}}$ would dramatically relax claimed constraints on 
light-WIMPs from XENON10 and XENON100 \cite{x10leff,xelast}.

An attempt has not been made here to simulate earlier measurements of  
$\mathcal{L}_{\text{eff}}$ as a function of neutron scattering angle. 
However, at least in the analysis of the experiments described in 
\cite{prevexp,manamana1,manamana2}, 
not only the overall normalization in rate was left as a free 
parameter (enabling the same deleterious effects illustrated here
by Fig.\ 4), 
but  additional degrees of freedom were introduced into the 
fits by way 
of a background model intended to account for 
neutron scattering in inert components, largely dominant in those 
other geometries$^{14}$.\footnotetext[14]{The detector used by Manzur {\it et 
al.} is properly designed for the job at hand, with only a bare minimum of 
inert parts around the LXe cell. The largely dominant ``materials background'' in 
\cite{manamana1,manamana2} arises from a suboptimal detector design, causing neutrons 
to have a very small probability 
of reaching and exiting the active cell without interacting on inert 
materials, creating a background that completely swamps the signals 
sought.} 
This is clearly a flawed approach to the analysis of such 
experimental data. The dismayingly poor quality of the calibration data in 
\cite{manamana1,manamana2} is worth emphasizing at this point: an interested reader is 
invited to inspect the discussion around Fig.\ 4-10 in 
\cite{manamana2} for crucial steps in the 
data analysis obviated in \cite{manamana1}. In the opinion of this 
author and others \cite{priv}, 
once the obvious residual threshold effect surviving the subtraction of 
accidentals is corrected for, there is little to no evidence in the 
dataset of \cite{manamana1,manamana2} for any usable calibration 
information at and below 10 keV$_{r}$. This calibration played a central role 
in the attempted justification of recent XENON100 limits \cite{xelast}.

\begin{figure}
\includegraphics[width=7.4cm]{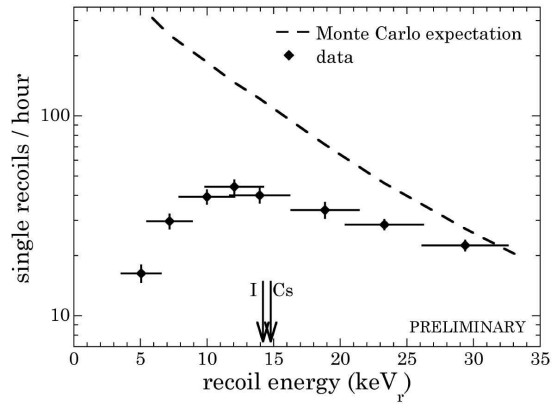}
\caption{Deficit in measured vs. expected single Cs,I recoil rate 
under irradiation of a CsI[Na] scintillator with 2.8 MeV 
monochromatic neutrons. Vertical arrows mark the value of the 
kinematic cutoff in recoil energy expected for 
ionization and scintillation in this medium (\cite{juandan}, see text).}
\end{figure}

To finalize, Fig.\ 6 is offered to illustrate the reality of the concerns 
expressed in this paper. It displays the rate of single Cs or I recoils in a 
CsI[Na] scintillator irradiated by a 2.8 MeV DD neutron generator in 
a setup similar to 
that used by Manzur {\it et al.} These data are part of an ongoing 
series of precision quenching factor calibrations for CsI[Na], CsI[Tl] and 
NaI[Tl] at the University of Chicago, employing combined electron 
recoil (Compton scattering) and nuclear recoil measurements on a 
single goniometric table. While these recent data should be considered 
preliminary, a large deficit in signal rate is observed, with an 
onset coincident 
with the predicted kinematic cutoff for Cs,I recoils in CsI[Na] \cite{juandan}. 
At the time of this writing, data acquisition triggering throughput 
and threshold efficiency effects are found to be far from sufficient to account for its 
magnitude (the measured low-energy light 
yield of this scintillator in this setup is $\sim$9 
PE/keV$_{ee}$). Following the incorrect 
analysis procedure criticized in this section, a monotonically increasing 
quenching factor towards zero recoil energy would be naively derived 
from the CsI[Na] data at hand. The lowest energy data point 
centered at 5 keV$_{r}$ would be assigned 
a quenching factor of $\sim$20\%, with the highest energy points in 
agreement with earlier measurements \cite{csina} at $\sim$8\%. 
With a correction in place for this 
systematic  deficit in rate, the derived quenching factor should be 
much 
smaller for the lowest energies measured. An upcoming 
publication will discuss the implications of these measurements for 
the exact location of the DAMA favored region in light-WIMP phase 
space \cite{danandco,juandan}, impact on efforts to reproduce the annual modulation effect using 
CsI[Tl] crystal arrays \cite{kims}, and prospects for a low-energy neutrino 
measurement using CsI[Na] at an intense neutron spallation source \cite{todd}.

\section{IV: Conclusions}

Even after considering the difficulties described in the 
introduction, common to all techniques, it is possible to distinguish 
between detector technologies more adapted to a search for 
light WIMPs 
than others, based exclusively on the existing knowledge of the response of 
the detector in the few keV recoil energy region. Two families of 
devices can be examined: those for which a reliable 
low-energy scale exists, and those for which it is presently 
lacking. Belonging to the first group, DAMA/LIBRA NaI(Tl) scintillators benefit from a convenient 
low-energy signal at 3.2 keV originating in a $^{40}$K 
internal contamination, able to anchor the energy scale near 
threshold, even if a relatively small uncertainty still remains in the quenching 
factors that define the recoil energy scale \cite{juandan,danandco}. 
CDMS germanium bolometers \cite{cdmslast} profit from peaks at 1.3 
keV and 10.4 keV
from $^{71}$Ge 
activation following periodic neutron calibrations. 
The quenching factor measured in CDMS-Ge is similar to theoretical 
expectations and the observations from
independent germanium experiments. CRESST bolometers are able to use 
narrow lines of known origin at 3.6 keV and 8 keV for a low-energy 
reference, having 
measured the quenching factors for the 
recoiling species in their crystals using a number of 
techniques \cite{cresst2}. CoGeNT detectors benefit from a 
number of known narrow cosmogenic lines in the range 1.1-11.1 keV and 
an optimal linearity and energy resolution.   
Their quenching factor has been measured in a dedicated reactor 
experiment down to sub-keV recoil energies, obtaining an 
excellent agreement with expectations and other measurements \cite{Ge2}. Belonging to 
the second group we can identify CDMS silicon bolometers: 
spatial corrections are known to affect their recoil energy 
scale, generating a large disagreement with theoretical values of the 
quenching factor (even larger when compared to
other experimental values), and a potentially large 
shift in
recoil energy \cite{danandco}. At the time of this 
writing, detectors based on LXe lag 
behind all other dark matter detection techniques in a knowledge of the 
energy scale and of the 
mechanisms governing signal production by few keV$_{r}$ nuclear recoils. 
Recent efforts to provide an electron-recoil low-energy calibration reference via a 
$^{83m}$Kr source \cite{kr83} are a welcome step in the right 
direction.

To summarize, important flaws in the methodology used to determine the value 
of $\mathcal{L}_{\text{eff}}$, $Q_{y}$ and the recoil energy scale in liquid xenon detectors 
have been described. Data-quality concerns affecting recent 
calibrations, 
such as those in \cite{manamana1,manamana2}, have been pointed out. 
Once these issues are addressed, 
an agreement with the expected decrease in ionization and scintillation yield 
from nuclear recoils below kinematic cutoff should be obtained for 
this medium. In particular, the 
traditionally observed deficit in low-energy response 
to AmBe neutron-induced recoils in LXe should be easier to understand 
by the inclusion of well-known radiation effects presently being ignored. 
A proper control of the systematics affecting 
monochromatic neutron scattering 
measurements of $\mathcal{L}_{\text{eff}}$ and $Q_{y}$ has been shown  
to be lacking: bold as this may sound, none of the measurements of 
this type performed
up until now below $\sim$10 keV$_{r}$ by the XENON10 or XENON100 collaborations can be 
assigned much worth. 
Separately, an effort should be made 
to evolve the ``best-fit Monte Carlo'' method to 
include radiation effects known to mediate 
the generation of signal carriers (free electrons, direct 
scintillation) 
{\it before} an attempt to match simulations and data is made. Proper 
account of the uncertainty in the recoil energy scale should be taken 
before attempting to extract limits from spectra generated by this method. In its 
present na\"{i}ve form, it is not possible to defend it. 

Markedly different values of $\mathcal{L}_{\text{eff}}$ and $Q_{y}$ should be 
expected from improved LXe calibration methodologies. 
The recoil energy scale of ionization (S2 light) spectra should also be 
affected by the proposed improvements, as described in the commentary around 
Figs.\ 1 and 2.

In view of the numerous 
issues raised here, one is forced to 
conclude that recent attempts to 
extract light-WIMP sensitivity from 
XENON10 and XENON100 data \cite{x10leff,xelast} are  
premature and  
overly optimistic. The sensitivity of present LXe detectors to WIMPs heavier 
than $\sim$10 GeV/c$^{2}$ is also 
affected by the considerations presented here, albeit to a lesser 
(but calculable) extent.

The author is indebted to D.N. McKinsey for many frank exchanges 
that made the present analysis possible, and for calling his 
attention to the limitations affecting the measurements in 
\cite{manamana1,manamana2}. D. Hooper, N. Weiner and K. Zurek provided useful 
suggestions.

\end{document}